# Spontaneous electric polarization and Magneto-dielectric coupling in ceramic multiferroic $Co_3TeO_6$

Harishchandra Singh[1], Haranath Ghosh[1], A. K. Sinha[1], M. N. Singh[1], G. Sharma[2], S. Patnaik[2], S. K. Deb[1]

[1]Indus Synchrotron Utilization Division, Raja Ramanna Center for Advanced Technology, Indore -452013, India.

[2]School of Physical Sciences, Jawaharlal Nehru University, New Delhi-110067, India.

Dated: September 24, 2013

*We report observation of magneto-electric and magneto-dielectric couplings in ceramic $Co_3TeO_6$. Temperature dependent DC magnetization and dielectric constant measurements together indicate coupling between magnetic order and electronic polarization. Strong anomaly in dielectric constant at ~ 18K in zero magnetic field indicates presence of spontaneous polarization. Observations like weak ferromagnetic order at lower temperature, field and temperature dependences of the ferroelectric transition provide experimental verification of the recent theoretical proposal by P. Toledano et al., Phys. Rev. B 85, 214439 (2012). We provide direct evidence of spin-phonon coupling as possible origin of magnetic order.*

Pacs: 77.84.-s, 75.85.+t, 77.80. –e

     Intriguing nature of strongly coupled magnetic and electronic degrees of freedoms in multiferroics provide challenges and opportunities to condensed matter physics from fundamental to applications (in spintronics, information storage devices, integrated circuits and so on [1]) based science. It appears from earlier understandings that only certain types of magnetic orders like helical, conical, spiral, frustrated spin couple to ferroelectricity [2] in two following ways. The electric polarization and the anomaly of dielectric constant can arise from atomic displacement−quanta of which, phonons in turn couples to spin [3]. The other possibility is that the electric polarization may arise from electronic wave function and thus the density distribution. It was predicted by Katsura, Nagaosa and Balatsky that the magnetoelectric effect can be induced by ``spin current'' [4]. The spin current and internal electric field couples in the form of Dzyaloshinskii-Moriya interaction $\mathbf{P} \propto \sum \mathbf{e}_{ij} \times (\mathbf{s}_i \times \mathbf{s}_j)$ where $\mathbf{e}_{ij}$ is the vector connecting two neighbouring spins $\mathbf{s}_i$ and $\mathbf{s}_j$ [5].

     Very recently $Co_3TeO_6$ (belonging to $A_3TeO_6$ family; A = Mn, Ni, Cu, Co) multiferroic shows interesting physical as well as low temperature magnetic properties, like thermal

variations of magnetization, susceptibility, specfic heat indicating very complex magnetic structures [6]. Neutron diffraction studies on $Co_3TeO_6$ (last but one of ref. [6]) clearly indicate first-order multi-*k* phase transitions, a sequence of three antiferromagnetic (AFM) phases accompanied by magnetoelectric effects. The incommensurate phase I emerging at $T_N$ = 26 K whereas the two commensurate phases at 21.1 K (phase II) and 17.4 K (phase III) respectively. The evolution of phase II from phase I take place through strong first order transition. Theoretical Landau free energy analysis taking care of irreducible representations and monoclinic magnetic group symmetry suggests a strong magnetoelastic effect may be involved [7]. This further suggests possible abrupt change in interatomic spacing causing the first order transition which in turn influence exchange energy from phase I to II. A phenomenological explanation of magnetoelectric behaviour of $Co_3TeO_6$ is also developed in ref. [8]. In contrast, there is smooth second order transition from phase- II to III indicating no significant discontinuity of lattice parameter. As theoretically explained [7] monoclinic symmetry of phase III (2´) permits spontaneous weak magnetization as well as spontaneous polarization which is a single spontaneous zero field polarization component $P_y$ [7]. While the former (weak magnetization) is consistent with domain structure in second harmonic generation (SHG) (last ref. of [6]) measurements, the spontaneous zero field polarization [7] is yet to be experimentally verified.

We present in this letter experimental evidence of all the three phases discussed above through magnetoelectric and magnetodielectric measurements in *ceramic* $Co_3TeO_6$ (CTO), synthesized using conventional solid state reaction route. As temperature is lowered towards 21.1 K our measured dielectric constant changes its trend showing a sharp upward turn forming a peaked structure at around 17.5 K at zero magnetic field. Monoclinic crystal structure with C2/c space group has been verified through Rietveld Refinement on the Synchrotron X-ray diffraction (SXRD) data. $Co^{2+}$ oxidation state and mixed coordination geometry of Co in CTO sample is confirmed by X-ray absorption near edge structure (XANES) measurements. A thorough and careful Rietveld refinement of low temperature (LT)SXRD data shows a sharp discontinuity in lattice parameters at around 21 K confirming a first order transition between phase I to II, the same is small in transition from phase II-III. Well inside phase III, our zero field dielectric constant shows very similar to $(T_2 - T)^2$ behaviour as predicted theoretically [7]. With magnetic field such behaviour changes with lowering of temperature, showing suppression in overall magnitude indicating coupling between magnetic order and polarization. Temperature dependencies of *difference* in various average Co-O bonds follow *similar* trend as that of the magnetization indicating direct coupling between the lattice and spin degrees of freedom. Furthermore, field dependence of $T_2$ (H) and weak hysteresis indicating spontaneous magnetization at 2K are also verified. Thus, our data provides pioneering experimental verification of the theoretical prediction of spontaneous zero magnetic field polarization and being consistent with measurements in all the other phases.

Ceramic cobalt tellurate are synthesized using conventional solid state reaction route. The reactants used are cobalt oxide $Co_3O_4$ (Alpha Easer 99.7 %) and tellurium dioxide $TeO_2$ (Alpha Easer 99.99 %). The reactants in stoichiometric ratio are first calcined (700°C for 10 hrs) and then recalcined (800°C for 20 hrs). SXRD (room temperature to 7 K) and XANES measurements are performed at angle dispersive x-ray diffraction (ADXRD) beamline (BL-12) [8] at Indus-2 synchrotron source. LTXRD are performed in a liquid helium cryostat (Advanced Research Systems Inc Model No. LT-3G) using image plate based MAR 345 dtb area detector. The temperature of the sample is stabilized within 0.1 K. SXRD patterns are corrected using $LaB_6$ and Si standards. The two dimensional patterns are integrated using the programme fit2D [9]. The refinements of the structural parameters from the difraction patterns are obtained using the Rietveld analysis employing the FULL-PROF program [10]. Magnetic measurements are carried out with a vibrating sample magnetometer in the magnetic property measurement system (MPMS-SQUID VSM) of Quantum Design, USA. The temperature stability at the sample chamber is better than ± 0.2 K. Temperature dependent dielectric measurement is performed using 1920 QUADTech LCR meter in Cryogenic CFM system. The sample used for the capacitance measurement is 0.4 mm thick and 78.5 $mm^2$ area.

CTO adopt monoclinic structure with space group C2/c as reported earlier for crushed single crystal as well as ceramic CTO. The experimental, calculated, and difference powder-diffraction profiles are shown in Figure 1. For the purpose of structural verification, a comparative table of lattice parameters of CTO sample is presented in table 1. The corresponding bond valance sum calculations [11], show the presence of $Co^{2+}$, $Te^{6+}$ and $O^{2-}$ ions, which are further confirmed using XANES and X-ray photo electron spectroscopy (XPS) studies (this along with detailed scanning electron microscopy (SEM), energy dispersive X-ray (EDX) studies will be reported elsewhere). Large variations in Co-O bond lengths (1.87Å to 2.93Å) as well as Co-O-Co bond angles ($90°$ to $140°$) have been found, in agreement with earlier reports [13].

TABLE. 1. Comparison of lattice parameters of ceramic CTO sample having monoclinic structure with space group C 2/c with the reported values of lattice parameters.

| Sr.No. | a | b | c | β | Ref. |
|---|---|---|---|---|---|
| 1 | 14.8167 | 8.8509 | 10.3631 | 94.90 | [11] |
| 2 | 14.8014 | 8.8379 | 10.3421 | 94.83 | Hudl et al.,[6] |
| 3 | 14.7526 | 8.8139 | 10.3117 | 94.905 | [12] |
| 4 | 14.8113 | 8.8394 | 10.3589 | 94.834 | [13] |
| 5 | **14.7856** | **8.8287** | **10.3315** | **94.839** | **Our results** |

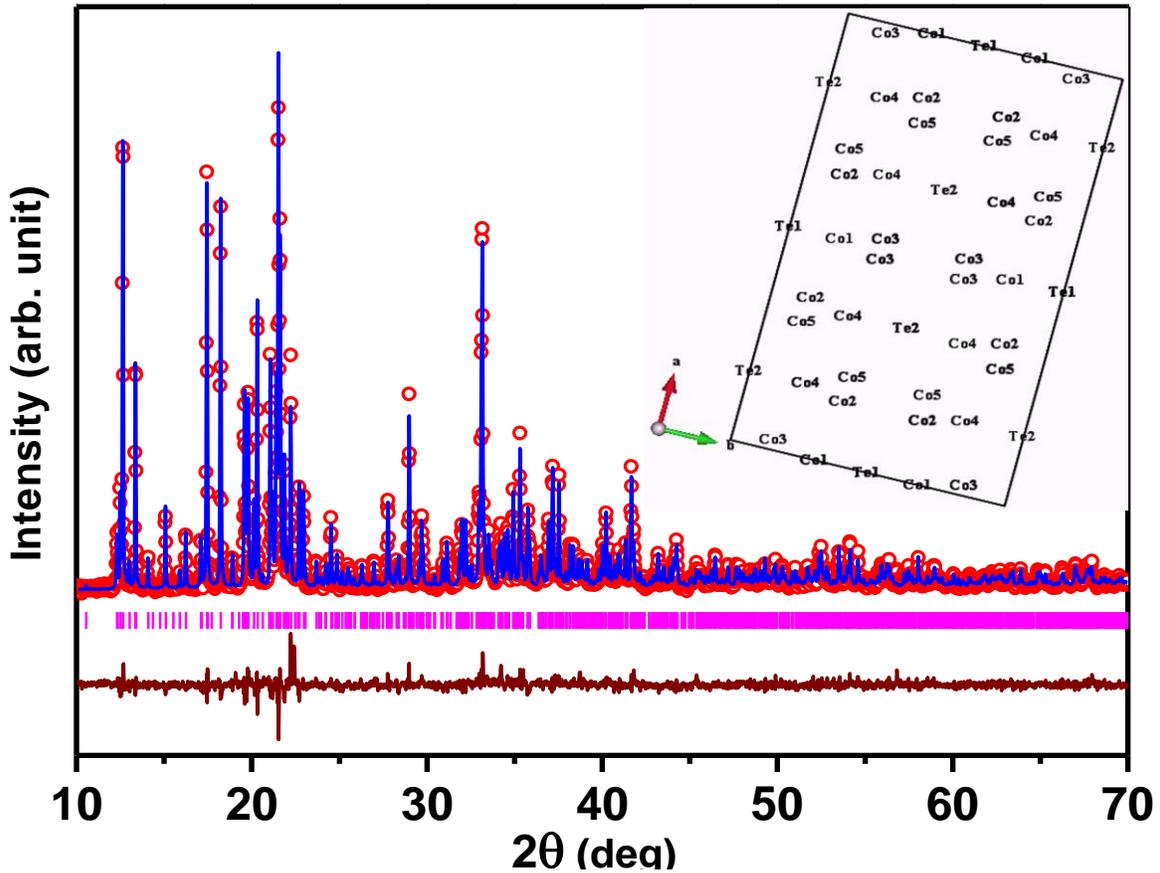

Figure 1. Synchrotron powder X-Ray diffraction patterns recorded at room temperature from CTO sample. Open circles represent the observed data, and solid line is the fit obtained by the Rietveld method using the monoclinic structure having space group C2/c. The vertical bar lines show Bragg positions. The line beneath the pattern records the difference patterns between the observed and calculated Intensity patterns. Model two dimensional view of the CTO crystal structure (Oxygens are not shown).

The CTO structure contains inequivalent five Co and two Te cations, where Co occupies 4e and 8f positions, whereas, Te occupies 4b and 8f. Te cations occupy octahedral sites whereas Co sits in tetra, penta and octahedral sites. The corresponding crystallographically distinct nine O atoms sit in 8f positions.

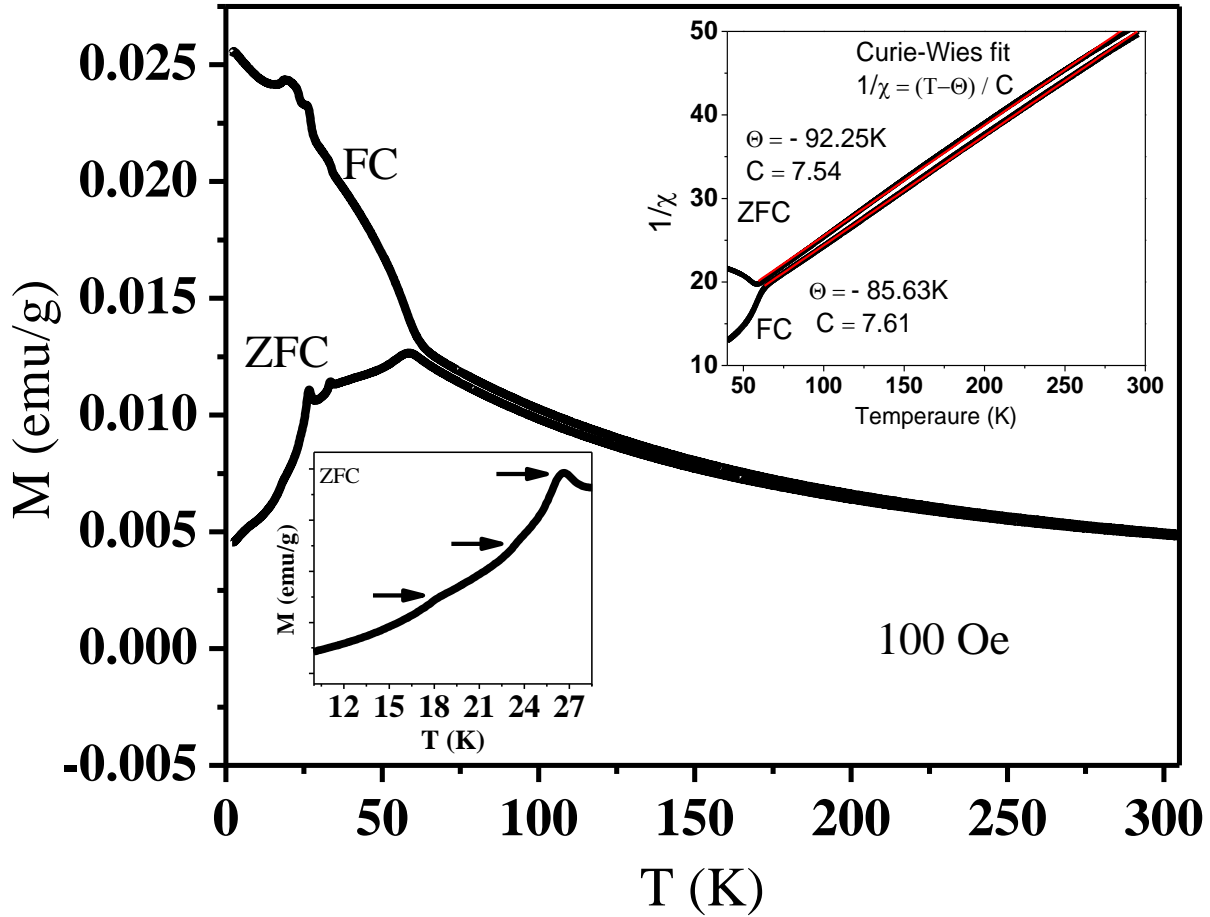

Figure 2. ZFC and FC Magnetization vs temperature measurement in the presence of 100 Oe magnetic field. Inset (right top corner) shows the Curie-Wies fit of the corresponding ZFC and FC mode. Inset (bottom left corner) indicates the occurence of three different magnetic transitions.

Temperature dependent DC magnetization measurements are performed in presence of constant applied magnetic field [shown in Fig. 2]. ZFC and FC measurements are performed in weak magnetic field of magnitude 100 Oe (weak enough in comparison to anisotropic field, around 400 Oe). We fitted the high temperature part of the $\chi^{-1}$ curve for both ZFC and FC (T > ~ 60K) to the Curie-Weiss law $\chi(T)^{-1} = (T - \Theta) / C$ where $\Theta$ is the paramagnetic Curie temperature and C is the Curie-Weiss constant. Parameters are presented in Fig. 2. The negative Curie-Weiss temperature indicates antiferromagnetic (AFM) order. The thermal irreversibility is seen in ZFC and FC magnetization data near 60K called blocking temperature. We observe three main magnetic transitions at around 26K, 21.5K and 17.5K in ZFC and FC magnetization curves (consistent with the neutron diffraction data [6]. Gradual decrease of magnetization (ZFC) below blocking temperature, generally indicate purely AFM transition. However, in our case, magnetization first started decreasing from ~ 60K up to ~ 26K and then decrease up to ~ 2K (see the inset Fig. 3). Different temperature dependencies of magnetization in the temperature ranges 26K to ~ 22 K, from 22 K to 17.5 K and below 17.5 K are observable (see inset Fig. 2), indicating presence of different phases observed in neutron scattering [6]. Around 35K an additional transition is observed. Such thermal behaviour of magnetization should be compared with that of the lattice parameters presented in Figs. 5, 6. Since there is no structural changes observed throughout the temperature ranges, it is conceivable that nature of phase transition from phase I to II is first order type. This is so because at around 22 K one observes sharp discontinuity in all the lattice parameters (shown in Figs. 5, 6).

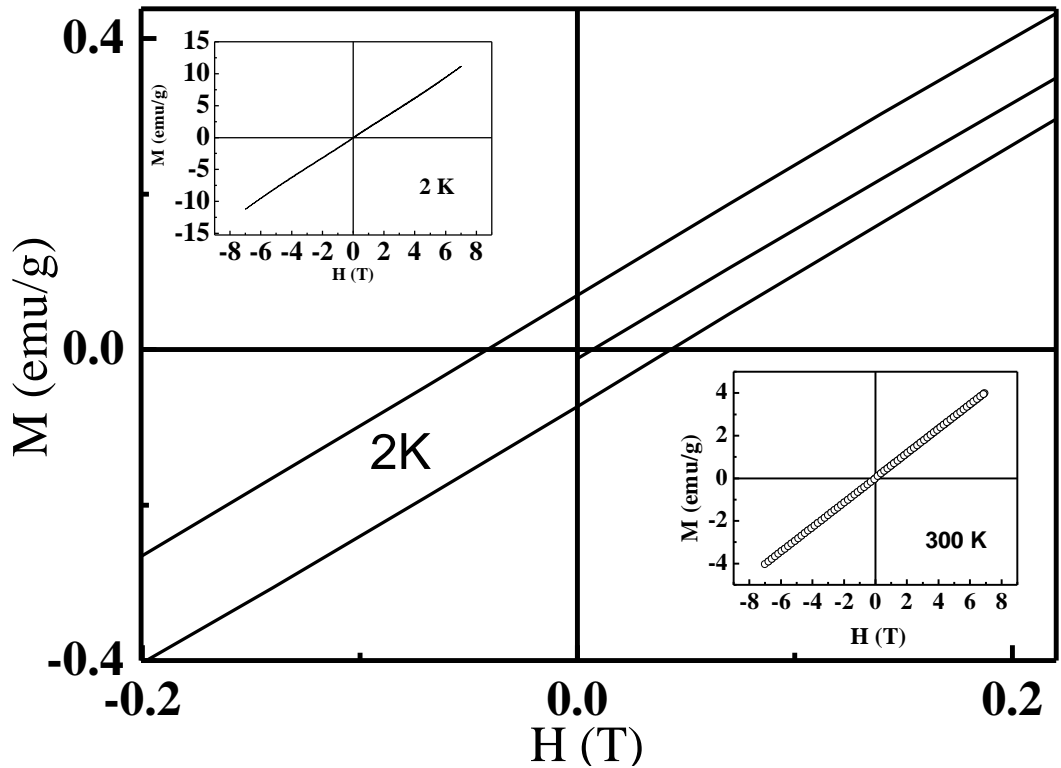

Figure 3. Magnetic hysteresis loop measurement at 2 K in FC mode and 300 K in FC confirm weak ferromagnetism at low temperature and paramagnetism at room temperature is shown in the figure.

Furthermore, a closer look at Fig. 6 which demonstrates the average Co-O bond distances of various Co s, indicates that at 16 K i.e. phase -III) the bond length changes very slowly, whereas at 25 K (phase -I) changes drastically. Discontinuities in all the lattice parameters are limited only to a small temperature range. Therefore, phase I in our ceramic sample is an incommensurate antiferromagnetic type whereas the phases II, III are commensurate type. This becomes more obvious from a closer view of Fig. 6, Co-O bond distances, both at room temperature (296 K) and 7K show nearly same behavior indicative of commensurate nature. In contrast, the bond distances at 25 K are exactly asymetric. Strong coupling between spin and lattice degrees of freedoms become evident from the Fig. 6 (b). A comparision with Fig. 2 shows overall similarity between the thermal variation of difference in average Co-O bond distances (a detailed study of which will be reported elsewhere [14]) to that of the magnetic moment (shown in Fig. 2).

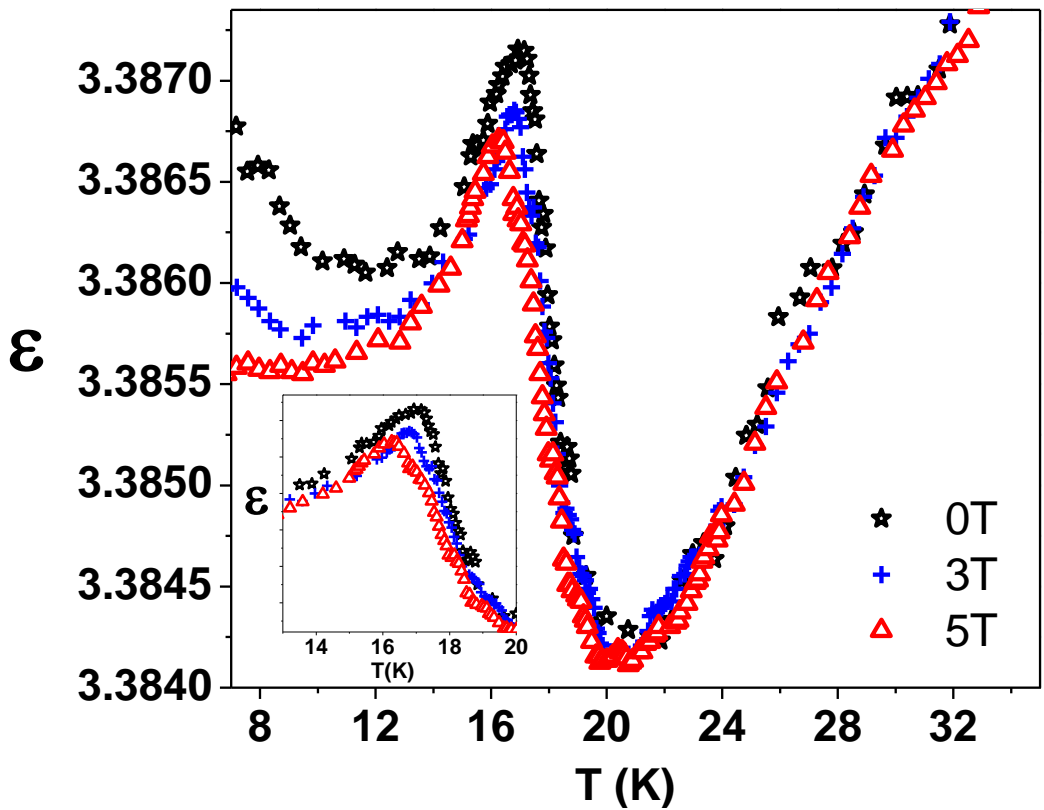

Figure 4. Dielectric constant vs temperature measurement with and without applied magnetic field. Behaviour of zero magnetic field dielectric constant near $T_2 = 17.5$ K indicates occurence of spontaneous electric polarization. The peak position moves down to lower temperatures as field is increased (see inset).

According to the prediction by Toledano et al., [7] spontaneous electric polarization is also accompanied by spontaneous magnetization in phase III. To confirm these, we have performed magnetic hysteresis loop measurements at 2K and 300K [Fig. 3] as well as dielectric constant measurements [Fig. 4]. Weak hysteresis loop at 2K confirms the presence of weak ferromagnetism in the sample. The dielectric constant [Fig. 4] shows a sharp upward turn at around 21 K which with further lowering in temperature at around $T_2 = T_C^{FE} \approx 17.5$ K, a strong peak in the dielectric constant is observed *in absence* of magnetic field (Fig. 4). Such a sharp upward turn in the dielectric constant appears exactly at the same temperature where all the lattice parameters show discontinuity. This shows spontaneous polarization in our sample indicating strong coupling between lattice degrees of freedom and spin. Such changes in lattice parameter is not causing any structural changes as far as space group is concerned, hence these modifications in lattice parameters may cause change in the exchange correlation (spin-phonon coupling) leading to incommensurate to commensurate transition with lowering in temperature, in agreement with neutron diffraction data [7].

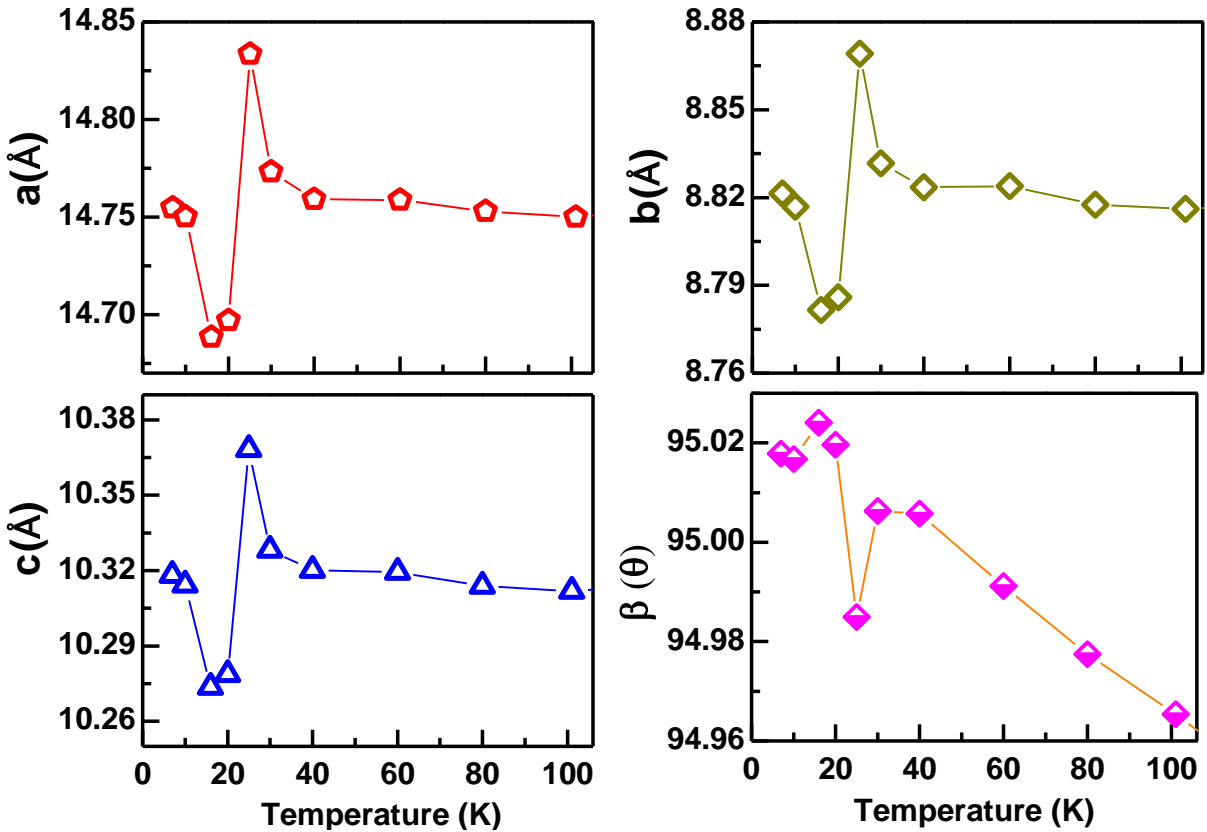

Figure 5. Anomalies in the lattice parameters with temperature at around transition temperature 21 K clearly indicates incommensurate nature of transition between phase I to phase II and posible magnetoelastic phenomenon in CTO sample.

Furthermore, the transition between magnetic orderings in phase I, II would not cause large change in entropy to show up a sharp discontinuity in specfic heat measurement, however such discontinuity [6] is observed exactly at the same temperature. This further indicates strong lattice-phonon coupling in CTO. Therefore we conclude that there is a spontaneous polarization in our ceramic CTO sample which is not observed so far and hence provides experimental verification of the theoretical predication by P. Toledano et al., [7]. Furthermore, according to the theoretical prediction [7], the spontaneous polarization would show a temperature dependence in the phase III as $(T_2 - T)^2$ which agrees well with our observation in Fig. 4. The transition temperature $T_2$ decreases with increasing applied magnetic field (see inset Fig. 4). This observation is also in accordance with the theoretical prediction by Toledano et al., [7]. Since the electrical polarization (see Fig. 4) occurs at the same temperature as the magnetic transition (phase boundary between II and III, see Fig. 2) where weak ferromagnetism is also observed (Fig. 3), the two orders electrical polarization and magnetism are coupled. This is further established from the similarities in the temperature dependencies between magnetic moment and *difference* in Co-O bond distances.

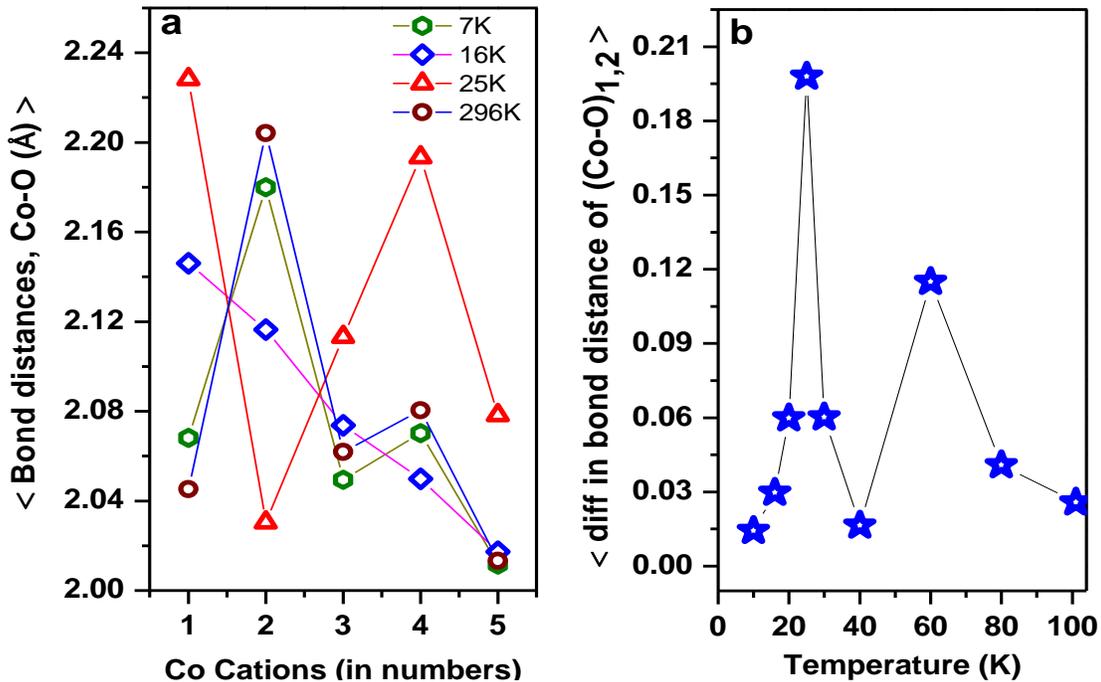

Figure 6. (a)Variation of average Co-O bond distances in $Co_3TeO_6$ at various temperatures. Nature of variations at very high and low temperatures maybe contrasted with that of 25 K. All the variations with respect to that of 16 K to be compared (b) Qualitative resemblence of thermal variation of magnetic moment (see Fig. 2) with that of difference in average Co-O bonds (between $Co_1$ and $Co_2$) -- a direct evidence of spin phonon coupling.

One of the recent issues in CTO is whether it exhibits zero magnetic field spontaneous polarization along a particular direction in the low temperature phase (below 18 K) or not. Whether magnetization also appears along other orthogonal directions or not. We presented partial answers to the above in this letter. Temperature dependent DC magnetization reveals complicated magnetic structure in CTO and signature of all the three phases observed in neutron diffraction. Incommensurate, commensurate magnetic transitions and direct evidence of spin-phonon coupling are obtained through anamolies in the temperature dependences of lattice parameters, average Co-O bond distances, and resemblance of magnetic moment with *difference* in average Co-O bond distances respectively. The measured dielectric constant exhibits a steep upward turn at around 21 K forming a well defined peaked structure at around 17.4 K ($T_2$). On further lowering in temperature, it behaves very similar to $(T_2 - T)^2$ and on applying magnetic field the $T_2(H)$ decreases with increasing field. On this phase (at lower temeprature) weak magnetization is developed as evidenced through hystersis loop. These observations are in very close agreement with the recent theoretical prediction [7] and may validate the same. We believe our results will lead to further experimental and theoretical studies in CTO.

*Acknowledgements:* We thank A. Sundaresan, B. Rajeswaran for magnetic measurements and S. M. Gupta for useful discussions.